\documentclass[onecolumn,authoryear]{els-mrw}

\usepackage{graphicx}

\usepackage{url}
\usepackage{comment}
\usepackage{color}
\usepackage{slashed}  

\usepackage{hyperref}
\hypersetup{
    colorlinks = true,
    citecolor  = blue,
    linkcolor  = blue,
    urlcolor  = blue
}

\usepackage{mathtools, braket}
\usepackage{amsmath,amsfonts,amssymb,amsthm}

\newcommand{\LambdaQCD}{\Lambda_{\rm QCD}}

\newcommand{\zmax}{z_{\rm max}}
\newcommand{\MSb}{\overline{\mathrm{MS}}}

\usepackage[most]{tcolorbox}
\usepackage{xcolor}

\newtcolorbox{keypoint}{
  colback=gray!15,   
  colframe=gray!60,  
  boxrule=0pt,       
  left=6pt,
  right=6pt,
  top=6pt,
  bottom=6pt
}

\title{EncyclopediaNP}

\begin{document}
\chapter{Generalized parton distributions from lattice QCD}
\author[1]{Krzysztof Cichy\footnote{Principal author of this review. Contact information: krzysztof.cichy@amu.edu.pl}}
\author[2]{Martha Constantinou}
\address[1]{\orgname{Adam Mickiewicz University}, \orgdiv{Faculty of Physics and Astronomy}, \orgaddress{ul.\ Uniwersytetu Poznańskiego 2, 61-614 Poznań, Poland} }
\address[2]{\orgname{Temple University}, \orgdiv{Department of Physics}, \orgaddress{Philadelphia, PA 19122 - 1801, USA}}
\maketitle

\begin{abstract}
This chapter gives an overview of the recent progress in extracting generalized parton distributions from lattice QCD. We briefly recall the theoretical principles of GPDs and explain the two most common lattice approaches, Ji's quasi-distributions and Radyushkin's pseudo-distributions. In the second part of the chapter, we review the lattice results obtained from these two frameworks. Finally, we offer a discussion of future prospects of lattice extractions of GPDs.
\end{abstract}

\section{Introduction} 
\label{sec:intro}
Hadrons are complex bound states whose internal structure and dynamics are governed by quantum chromodynamics (QCD), a vital ingredient of the Standard Model that describes all fundamental particles and interactions. The naive constituent quark model describes baryons, such as the nucleon, as systems of three valence quarks (e.g., $uud$ for the proton or $udd$ for the neutron), and mesons, such as the pion, as a quark-antiquark pair (combination of the light $u$ and $d$ quarks and antiquarks, depending on the electric charge of the pion).
However, their actual physical state is a dense, non-perturbative combination of these valence quarks, gluons, and virtual quark-antiquark pairs (including ones of heavier flavors, like strange and charm), elementary color-charged particles collectively called partons. 
Due to the complexity of QCD, resolving how these fundamental degrees of freedom share the hadron's total momentum, charge, and spin remains a central challenge of modern nuclear and particle physics.

The first glimpse into nucleon's structure was accomplished back in the 1960s in seminal deep inelastic scattering (DIS) experiments at Stanford, see another chapter of this Encyclopedia on ``Deeply inelastic scattering (DIS) experiments -- overview''. These experiments played a similar role for the nucleon as the famous Rutherford's experiments for the atomic nucleus, i.e., they revealed that the nucleon is a composite object, hence one that possesses an internal structure.
DIS and similar processes are inclusive, meaning that only some of the scattering products are measured and, in particular, the initial hadron's fate is unknown.
From the theoretical point of view, the description of scattering involving composite hadrons is described through collinear factorization by separating long-distance and short-distance contributions, see another chapter of this Encyclopedia on ``QCD (collinear) factorization and the renormalization group''. The total scattering cross section is a convolution of
parton distribution functions (PDFs), encoding the non-perturbative part, and a partonic cross section, calculable in perturbation theory.
As such, PDFs are indispensable for analysis of collider experiments, e.g., in the Large Hadron Collider (LHC).
For a review of PDFs, see, e.g., Refs.~\cite{Gao:2017yyd,Kovarik:2019xvh,Ethier:2020way} and other chapters of this Encyclopedia on ``Parton distributions of the proton and precision QCD'', ``Global analysis of PDFs'', and ``Global analysis of helicity-dependent parton distribution functions''.

Despite the importance of PDFs, they only provide longitudinal momentum information, leaving aside any dynamics in the transverse plane and hence, omitting essential mechanical and structural properties of hadrons.
From the point of view of particle scattering, PDFs are not capable of describing exclusive processes, for which the initial and final hadronic states are fully specified and the scattering dynamics involves a momentum transfer to the hadron.
Thus, in the 1990s, the need to extend the PDF picture and description of hadrons was realized \cite{Mueller:1998fv, Ji:1996ek,Ji:1996nm,Radyushkin:1996nd}, introducing a new class of structural functions, generalized parton distributions (GPDs).
GPDs are powerful tools that unify and enhance the information encoded in PDFs, elastic form factors (EFFs) and generalized form factors (GFFs).
PDFs quantify the probability of finding a quark or gluon carrying a specific fraction of the hadron's longitudinal momentum, denoted by $x$.
EFFs are related to the spatial distribution of electric charge and magnetism in a hadron, measuring how it deflects an external photon, which transfers momentum to the hadron.
This momentum transfer is described by the Mandelstam variable, $t$.
GFFs, in turn, probe further aspects of the hadron's structure, e.g.\ gravitational form factors are linked to ``mechanical'' properties, via their relation to the energy-momentum tensor \cite{Polyakov:2002wz,Polyakov:2002yz,Polyakov:2018zvc}. 
GPDs generalize PDFs, EFFs and GFFs in the sense that the latter are certain limits of GPDs, PDFs being their forward limits ($t\rightarrow0$) and form factors being their Mellin moments (integrals of powers of $x$ with the GPD).
However, not all GPDs have a PDF limit -- thus, GPDs are more powerful objects allowing for a more comprehensive description.
In particular, they naturally lead to hadron ``tomography'' \cite{Burkardt:2000za,Ralston:2001xs,Diehl:2002he,Burkardt:2002hr}, a three-dimensional imaging of the internal structure of a hadron.
Moreover, GPDs can be related to the spin structure via Ji's angular momentum sum rule~\cite{Ji:1996ek}, which isolates the intrinsic spin and orbital angular momentum contributions.
In general, the $x$ dependence of GPDs includes two regimes with different partonic interpretations and distinct evolution equations, depending on the additional skewness variable, $\xi$, which accounts for the longitudinal part of the momentum transfer.
For $|x|>|\xi|$, GPDs probe the standard Dokshitzer-Gribov-Lipatov-Altarelli-Parisi (DGLAP) regime \cite{Dokshitzer:1977sg,Gribov:1972ri,Altarelli:1977zs}, known from the PDF case.
In turn, if $|x|<|\xi|$, a genuinely distinct Efremov-Radyushkin-Brodsky-Lepage (ERBL) region \cite{Efremov:1979qk,Lepage:1980fj} emerges.
Experimentally, GPDs are linked to hard exclusive processes, like deeply virtual Compton scattering (DVCS) \cite{Mueller:1998fv,Ji:1996nm,Radyushkin:1996nd,Collins:1998be} and deeply virtual meson production (DVMP) \cite{Radyushkin:1996ru,Collins:1996fb,Mankiewicz:1997uy}, among many others, see another chapter of this Encyclopedia on ``Exclusive processes in QCD''.
For reviews and introductions to various aspects of GPDs, see, e.g., Refs.~ \cite{Ji:1998pc,Radyushkin:2000uy,Diehl:2003ny,Belitsky:2005qn,Boffi:2007yc,Guidal:2013rya,Mezrag:2022pqk,Lorce:2025aqp} and other chapters of this Encyclopedia on ``Generalized parton distributions (GPDs)'', ``Spin Structure of the Nucleon: Overview'', ``GPD phenomenology'', and ``Nucleon and nuclear form factors: theory and experiment''.

However, extracting GPDs from experimental data is a notoriously difficult problem. This immense difficulty stems from a variety of reasons.
The primary theoretical hurdle is that GPDs cannot be measured directly. 
Instead, experimental observables give access to Compton Form Factors (CFFs), which are convolutions of GPDs with hard-scattering kernels, with the longitudinal momentum fraction $x$ integrated over.
As such, CFFs provide information only close to the $x=\xi$ line.
This so-called deconvolution problem gives rise also to the issue of ``shadow'' GPDs \cite{Bertone:2021yyz,Moffat:2023svr}, i.e., the property that infinitely many different GPDs can yield the exact same CFFs upon integration. 
Moreover, the sought for GPDs signal interferes with strong background, such as the Bethe-Heitler process \cite{Belitsky:2001ns}.
Additionally, multiple types of GPDs contribute to the same scattering cross-section at the same time.
For example, the description of a scattering for a spin-1/2 nucleon at leading twist requires four GPDs, both unpolarized and longitudinally polarized, leading to the necessity of difficult spin-asymmetry measurements with polarized beams, see, e.g., Ref.~\cite{Guidal:2013rya}.

Despite the fact that many of the above mentioned difficulties are expected to be alleviated in next years, in particular after the launch of the Electron-Ion Collider (EIC) at Brookhaven \cite{NAP25171,AbdulKhalek:2021gbh}, it is strongly desirable to exploit alternative paths of exploring GPDs.
Such a path is lattice QCD (LQCD), which provides a first-principles approach for exploring hadronic structure directly from the QCD Lagrangian.
In this framework, the QCD quark and gluon fields are put on a spacetime grid, representing a Euclidean spacetime.
This regularizes the infinite-dimensional QCD path integral and makes it amenable to large-scale numerical simulations on supercomputers.
Lattice QCD has a long history of investigating hadron structure, including observables related to GPDs.
For many years, the standard method was to access Mellin moments of GPDs via local matrix elements, see, e.g., Refs.~\cite{Hagler:2003jd,Gockeler:2003jfa,Alexandrou:2006ru,LHPC:2007blg,QCDSF:2007ifr,Alexandrou:2011db,Alexandrou:2013joa,Shintani:2018ozy,Bali:2018zgl,Alexandrou:2018sjm,Alexandrou:2019olr,Alexandrou:2019ali,Alexandrou:2020okk,Alexandrou:2021wzv,Alexandrou:2023qbg,Hackett:2023rif,Alexandrou:2025vto,Alexandrou:2026tjs}.
Mellin moments can be obtained from such local matrix elements via the operator product expansion (OPE).
This OPE-based program has been successfully realized on the lattice for many years, leading to several important insights on hadron structure, although without access to the $x$ dependence of partonic distributions.
In principle, such distributions can be reconstructed from a tower of moments. 
However, in practice, access is possible only to the lowest two or three moments, due to unfavorable signal-to-noise ratio for the higher ones and, even more importantly, to inevitable mixings with lower-dimensional operators. 
For more information about this approach from the lattice, see the recent review~\cite{Alexandrou:2026soz}, the FLAG report \cite{FlavourLatticeAveragingGroupFLAG:2024oxs}, and another chapter of this Encyclopedia on ``Lattice QCD for nucleon form factors''.

The Euclidean metric of spacetime obstructs direct access to partonic distributions, defined on the light cone.
Nevertheless, LQCD offers indirect access to the $x$ dependence of GPDs, via factorization of appropriate observables.
This thread is relatively recent, dating back to seminal papers of Ji \cite{Ji:2013dva,Ji:2014gla} in 2013-2014, which introduced the concept of quasi-distributions, objects based on non-local matrix elements of hadrons at a finite boost.
These objects can be matched to the desired partonic distributions in the framework of Large-Momentum Effective Theory (LaMET), which treats the finite but large hadron momentum as a physical scale that controls a systematic, effective field theory power expansion.
In 2017, Radyushkin introduced a related approach of pseudo-distributions \cite{Radyushkin:2017cyf}, based on the same lattice matrix elements, but instead of being based on a momentum expansion, the Euclidean observables are translated to their light-cone counterparts already in coordinate space, utilizing short-distance factorization (SDF), also called short-distance expansion (SDE).

The two above approaches have been studied very intensively theoretically and have been most widely used by the lattice community in practical studies. The remarkable progress achieved through these methods over the last few years forms the core focus of this chapter.
We also note that the chapter is restricted to quark GPDs, with gluon GPDs yet to be addressed by a lattice calculation.
However, it should be noted that alternative approaches exist.
The breakthrough sparked by Ji's work led to proposals of alternative approaches, some of which being revivals of earlier ideas \cite{Liu:1993cv,Detmold:2005gg,Braun:2007wv,Chambers:2017dov,Ma:2017pxb}.
Most of these proposals were put to use in the context of PDFs and distribution amplitudes, but all of them can, in principle, be utilized for GPD-related work.
In particular, the Compton amplitude approach for GPDs was explored in Refs.~\cite{CSSMQCDSFUKQCD:2021lkf,Hannaford-Gunn:2024aix} and the Euclidean hadronic tensor method in Ref.~\cite{Liang:2023uai}.
The former relies on computing the physical off-forward Compton amplitude from the lattice via the Feynman-Hellmann theorem, allowing the extraction of GPDs from a short-distance expansion of the scattering amplitude.
In turn, the latter calculates a four-point correlation of two currents on the lattice, leading to an Euclidean version of the hadronic tensor that can be
converted into its physical (Minkowski) counterpart by an inverse Laplace transform, giving access to, e.g., EFFs.
For general and topical reviews of lattice partonic distributions, see, e.g., Refs.~\cite{Lin:2017snn,Cichy:2018mum,Radyushkin:2019mye,Ji:2020ect,Constantinou:2020pek,Cichy:2021lih,Cichy:2021ewm,Constantinou:2022yye,Burkert:2022hjz,Lin:2025hka,Boer:2025ixc}.
The recent progress is also reported in another chapter of this Encyclopedia on ``Parton distribution functions from lattice QCD'', which is complementary to the present chapter.

The remainder of this chapter consists of two main sections and a summary. In Sec.~\ref{sec:framework}, we discuss the framework to extract GPDs from LQCD, starting with a short recollection of the theoretical principles of GPDs and continuing with the concepts that underlie the lattice calculation of bare matrix elements and the steps required to extract physical information on GPDs from them.
Section \ref{sec:results}, in turn, presents a brief review of lattice results obtained so far, using both LaMET and SDF.

\section{Framework to extract GPDs from lattice QCD}
\label{sec:framework}
\subsection{Theoretical principles}
\vspace*{1mm}\noindent\textbf{GPD correlators}\\
Generalized parton distributions are related to Fourier transforms of certain matrix elements that probe light-front correlations.
For a generic 4-vector $V$, with Cartesian coordinates $V=(V^0,V^1,V^2,V^3)$, one can define light-cone (Dirac) coordinates such that $V=(V^+,V^-,V_\perp)$, where $V^\pm=(V^0\pm V^3)/\sqrt{2}$ combine the temporal and spatial coordinates, while $V_\perp$ is the transverse spatial vector.
The generic definition of a quark GPD correlator reads:
\begin{align}
F_\Gamma^f(x,P,\Delta)=\int \frac{dz^-}{4\pi} \, e^{i x P^+ z^-} \left\langle P_f \left| \bar{\psi}^f(-z^-/2) \Gamma \mathcal{W}(-z^-/2, z^-/2) \psi^f(z^-/2) \right| P_i \right\rangle,
\label{eq:GPD}
\end{align}
where:\\
$F_\Gamma^f$ -- GPD correlator for Dirac structure $\Gamma$ that determines the GPD type (see below) and parton flavor $f$,\\
$x=k^+/P^+$ -- quark's longitudinal momentum fraction, where $k$ is the quark 4-momentum, $x\in[-1,1],$\\ 
$P_i, P_f, P$ -- initial, final and average hadron momentum 4-vectors ($P=(P_i+P_f)/2$),\\
$\Delta$ -- momentum transfer 4-vector ($\Delta=P_f-P_i$),\\
$\psi^f/\bar{\psi}^f$ -- flavor-$f$ quark/antiquark fields,\\
$\mathcal{W}(-z^-/2, z^-/2)$ -- Wilson line along the minus light-front direction (ensuring gauge invariance).\vspace*{2mm}\\
It is customary to define additional kinematic variables:\\
$\xi=-\Delta^+/2P^+$ -- the skewness variable, representing the longitudinal part of momentum transfer,\\
$t=\Delta^2$ -- invariant momentum transfer (one of Mandelstam variables).\vspace*{2mm}\\
With the above, one often writes the arguments of GPD correlators as $F_\Gamma^f(x,\xi,t)$.
In addition, the standard formulation of GPDs assumes vanishing average transverse momentum transfer, i.e., $P=(P^+,P^-,0_\perp)$, which leads to $P^2=2P^-P^+=M^2+t/4$, with $M$ the hadron mass.
Such a frame of reference, convenient in phenomenology, is referred to as the symmetric frame or the Breit frame.
Below, we will see that the symmetric frame has a significant drawback from the lattice perspective and we will discuss the crucial development of asymmetric frames of reference.

\vspace*{1mm}\noindent\textbf{GPD types}\\
At leading twist, in a spin-$S$ hadron, there are $(2S+1)^2$ chiral-even and $(2S+1)^2$ chiral-odd GPDs, $G^f$, for each quark flavor, defined using different Dirac structures employed in GPD correlators $F_\Gamma^f$. For $S=1/2$ (e.g., the nucleon):
\begin{itemize}
\item $\Gamma=\gamma^+$: $G=H,E$ -- unpolarized GPDs (chiral-even),
\item $\Gamma=\gamma^+\gamma^5$: $G=\tilde H,\tilde E$ -- helicity GPDs (chiral-even),
\item $\Gamma=i\sigma^{j+}\gamma^5$ ($j=1,2$): $G=H_T,E_T,\tilde H_T,\tilde E_T$ -- transversity GPDs (chiral-odd).
\end{itemize}
For example, $F_{\gamma^+}^f$ for a spin-1/2 baryon can be parametrized as
\begin{align}
\label{eq:unpolGPD}
F_{\gamma^+}^f(x, \xi, t) = \bar{u}(P_f) \left[ \gamma^+ H^f(x, \xi, t) + \frac{i \sigma^{+\mu} \Delta_\mu}{2M} E^f(x, \xi, t) \right] u(P_i),
\end{align}
where $u/\bar{u}$ are Dirac spinors for the initial/final hadron, and $\sigma^{+\mu} = \frac{i}{2} [\gamma^+, \gamma^\mu]$. 
These GPDs describe different possible polarizations of the quarks and the hadron and are, thus, related to different experimental situations. From the lattice perspective, access to all of them is possible in a similar way, as discussed below.
Beyond the leading twist, there are twice more twist-3 GPDs (involving a transverse index in $\Gamma$, e.g., $\gamma^{1,2}$ in the vector channel) and the number of twist-4 GPDs is the same as at twist-2 (e.g., $\gamma^-$ in the vector case), in all cases half are chiral-even and half are chiral-odd.
In principle, all GPDs can be accessed on the lattice upon some choice of the matrix $\Gamma$.

\vspace*{1mm}\noindent\textbf{Relation of GPDs to PDFs and form factors}\\
GPDs are three-dimensional objects, but they are related to simpler one-dimensional functions. In particular, the forward limits ($t\rightarrow0,\,\xi\rightarrow0)$ of the GPDs $H$, $\tilde H$ and $H_T$ are, respectively, the unpolarized, helicity and transversity PDFs, $f_1$, $g_1$, and $h_1$. Other types of leading-twist GPDs may also possess a non-vanishing forward limit, but there are no corresponding PDFs.
Similarly at higher twist, only some of the GPDs possess a PDF limit.
The other essential limit of GPDs is the one obtained upon integrating them with powers of $x$, corresponding to Mellin moments and various kinds of form factors. For example, the first moment of the unpolarized GPDs $H$ and $E$ (i.e., zeroth power of $x$) yields the Dirac and Pauli form factors, $F_1(t)$ and $F_2(t)$.
The second moments of these GPDs are related to gravitational form factors via a parametrization of the energy-momentum tensor.
We refer to other articles in this Encyclopedia and to several reviews for more discussions on the properties and uses of GPDs, see references given in Sec.~\ref{sec:intro}.

\subsection{Basics of GPDs from the lattice -- coordinate space}
The above definition of GPDs cannot be employed on the lattice, as it probes correlations along one of the light-front directions. This direction, obviously, requires access to the real time, while the time in lattice simulations is Wick-rotated, implying a Euclidean spacetime metric.
As a matter of fact, this was the stumbling block for lattice studies of PDFs and GPDs for a long time, restricting the practical work to extractions of Mellin moments, without addressing the $x$ dependence, see the short discussion in Sec.~\ref{sec:intro} and references therein.

\vspace*{1mm}\noindent\textbf{GPDs from spatial correlations}\\
In parallel, an independent thread emerged to access partonic distributions via non-local matrix elements (MEs) that probe purely spatial correlations, readily accessible on the lattice.
The generic form of such an ME reads\footnote{In the remainder of this chapter, we adopt the convention that Euclidean vectors have components written with lower indices.}
\begin{align}
\label{eq:latME}
F_\Gamma^f(z, P, \Delta) = \langle P_f|\,\bar{\psi}^f(-z/2)\,\Gamma\, \mathcal{W}(-z/2,z/2)\,\psi^f(z/2)\,|P_i\rangle.
\end{align}
It is, obviously, the integrand of the GPD correlator of Eq.~\eqref{eq:GPD}, but written along the (conventionally) third spatial direction and evaluated in Euclidean spacetime.
The ME is a function of the length of the Wilson line, $z$, and of the momentum variables.
The Dirac structure $\Gamma$ determines the GPD type:
\begin{itemize}
\item $\Gamma=\gamma_k$: $G=H,E$ -- unpolarized GPDs,
\item $\Gamma=\gamma_k\gamma_5$: $G=\tilde H,\tilde E$ -- helicity GPDs,
\item $\Gamma=i\sigma_{jk}\gamma_5$ ($j=1,2$): $G=H_T,E_T,\tilde H_T,\tilde E_T$ -- transversity GPDs,
\end{itemize}
for leading-twist, where the index $k$, replacing $+$ of the light-front GPD correlator, can be either 0 or 3.
The different MEs, $F_\Gamma$, can be decomposed into the GPDs, $G$, via parametrizations like Eq.~\eqref{eq:unpolGPD} and accessed by using different parity projectors, $\mathcal{P}_0=(1+\gamma_0)/4$ (unpolarized projector) and $\mathcal{P}_\kappa=(1+\gamma_0) i \gamma_5 \gamma_\kappa/4$ (polarized projectors, $\kappa=1,2,3$).
Concerning the gamma matrix indices $k$, both $k=0$ and $k=3$ can be used, in principle, but symmetry properties of practically used lattice actions lead to the preference of $k=0$ ($k=3$) in the unpolarized (polarized) case.
This is related to mixing with other Dirac structures induced by chiral symmetry breaking in the lattice action for the commonly used discretizations, such as Wilson-type fermions \cite{Constantinou:2017sej}.
For example, the use of $\gamma_3$ for unpolarized PDFs/GPDs implies undesired mixing with the scalar or the pseudoscalar structures, which is avoided upon using $\gamma_0$.
As the final remark, we note that to make connection to light-front correlations, the average momentum in the direction of the Wilson line, $P_3$, needs to be as large as possible.
The connection becomes, in fact, exact in the $P_3\rightarrow\infty$ limit, i.e., in the infinite momentum frame.
Meanwhile, at finite $P_3$, the connection can be made via perturbation theory, see below.

\vspace*{1mm}\noindent\textbf{Large Momentum Effective Theory and Short-Distance Factorization}\\
The approach based on the above introduced MEs originated in 2013, from two seminal papers by Ji \cite{Ji:2013dva,Ji:2014gla}, which discovered a systematic connection between these Euclidean MEs, evaluated for a momentum-boosted hadron, and the light-front (Minkowski) MEs.
It is known as the large momentum effective theory (LaMET) and the Fourier transform (FT) of MEs of Eq.~\eqref{eq:latME} yields the so-called quasi-distributions.
It was shown by Ji that quasi-distributions and their light-front counterparts have the same infrared (IR) structure and differ only in the ultraviolet (UV) regime.
Hence, this difference can be evaluated perturbatively and subtracted from the quasi-distribution.
Formally, there is a factorization relation between the Euclidean quasi-function and the light-front distribution, valid up to power-suppressed effects of $\mathcal{O}(\LambdaQCD^2/P_3^2)$, i.e., higher-twist effects (HTEs).
The above factorization, pertaining to quasi-distributions, can also be performed before the FT, i.e., in coordinate space (crucially, based on the same lattice input).
This was noted by Radyushkin \cite{Radyushkin:2017cyf,Radyushkin:2019mye} in 2017 and is currently known as short-distance factorization (SDF) or pseudo-distributions.
It should be emphasized that the two approaches, LaMET and SDF, are equivalent in the infinite momentum limit, but the different order of FT and factorization implies practical differences in systematic effects at finite $P_3$ and different form of power corrections, of $\mathcal{O}(z^2\LambdaQCD^2)$ in SDF.
As a matter of fact, these differences can be used as an advantage in proposed combinations of LaMET and SDF \cite{Ji:2022ezo}, which is further discussed below.

It is beyond the scope of this chapter to discuss in detail all of the aspects of LaMET and SDF, for which we refer to a dedicated chapter of this Encyclopedia on ``Large Momentum Effective Field Theory'' and several reviews \cite{Lin:2017snn,Cichy:2018mum,Ji:2020ect,Constantinou:2020hdm,Constantinou:2020pek,Cichy:2021lih,Cichy:2021ewm,Constantinou:2022yye,Lin:2025hka}, which also contain a selection of results obtained with these methods.
Below, we outline the key elements of quasi-/pseudo-distributions, for a self-contained presentation.
We start with brief general discussions and illustrate applications of several of these elements in the next section, reviewing the recent practical work.

\vspace*{1mm}\noindent\textbf{Extraction of bare MEs}\\
The first task of a GPD calculation consists in choosing lattice simulation parameters, such as the bare coupling, the lattice volume and the quark masses.
In practice, the choice is limited by the available gauge field configuration ensembles, typically generated by large lattice collaborations.
Assuming an ensemble has been chosen, the workflow for GPDs first includes evaluating bare MEs of Eq.~\eqref{eq:latME}.
They are obtained from a suitable ratio of three-point and two-point functions, $C^{\rm 3pt}$ and $C^{\rm 2pt}$.
Both are constructed by employing a hadron interpolating operator at some source position to create a desired hadron and again at some sink position to annihilate it.
In the case of $C^{\rm 3pt}$, there is additionally an insertion operator that probes the internal hadronic structure.

It is worth to emphasize that the calculation of the three-point function is the most computationally expensive part of the simulation. The two-point function is usually obtained from a solution of the discretized Dirac equation on a point source, namely $D\psi=\eta$, where $D$ is the Dirac operator and $\eta$ is a vector with only one non-zero entry. Utilizing the Wick's theorem, $C^{\rm 2pt}$ can be constructed from the so-called point-to-all propagator, $\psi=D^{-1}\eta$, the vector that solves the Dirac equation and describes the quark propagation from the source to all other lattice sites. Meanwhile, the presence of the insertion operator in $C^{\rm 3pt}$ implies that one needs an all-to-all propagator, i.e., in principle, the full inverse of $D$. Obviously, such a way of obtaining the all-to-all propagator would be prohibitively expensive and most modern computations employ the sequential method \cite{Martinelli:1988rr}, usually with a fixed sink. The fixed-sink sequential method consists in fixing the sink parameters, in particular the sink momentum $P_f$, and uses the point-to-all (``forward'') propagator to construct a ``sequential source''. This sequential source is then used in a new solution of the Dirac operator, yielding a ``backward'' propagator. Gluing the forward and backward propagators, together with the insertion operator, one obtains $C^{\rm 3pt}$. Importantly, the details of the insertion, as well as the source momentum $P_i$, can be decided after the Dirac operator inversions. This gives full flexibility for the type of GPDs that one can access.

The typically used ratio for GPDs reads:
\begin{equation}
R_{\Gamma}^f({\cal P}_{\kappa};\mathbf{P_f},\mathbf{P_i};t_s,\tau) = \frac{C_{\Gamma}^{\rm 3pt}({\cal P}_\kappa;\mathbf{P_f},\mathbf{P_i};t_s,\tau\
)}{C^{\rm 2pt}({\cal P}_0;\mathbf{P_f};t_s)}  
 \sqrt{\frac{C^{\rm 2pt}({\cal P}_0;\mathbf{P_i};t_s-\tau) C^{\rm 2pt}({\cal P}_0;\mathbf{P_f};\tau) C^{\rm 2pt}({\cal P}_0;\mathbf{P_f};t_s)}{C^{\rm 2pt}({\cal P}_0;\mathbf{P_f};t_s-\tau) C^{\rm 2pt}({\cal P}_0;\mathbf{P_i};\tau) C^{\rm 2pt}({\cal P}_0;\mathbf{P_i};t_s)}}\,,
\label{eq:ratio}
\end{equation}
where we have indicated the relevant variables of each function:
\begin{itemize}
\item the parity projector that is used; for the three-point function, there are 4 choices of $\kappa$, leading to the possibility of obtaining 4 MEs for each gamma matrix, which is crucial in the decomposition of MEs into (coordinate-space) GPDs,
\item the momentum variables; one requires two-point functions of both the source and the sink momenta, while the three-point function depends on both; the momenta are imposed to the functions by appropriate phase factors, e.g. $\exp(-i\mathbf{P}\cdot\mathbf{x})$ for the two-point function with momentum $\mathbf{P}$, and summation over all lattice sites,
\item the time variables; two-point functions depend on the temporal distance between the creation and the annihilation of the particle; the three-point function is usually obtained at a fixed source-sink temporal separation, $t_s$, and the structure-probing insertion operator is inserted at times $\tau=0,\ldots,t_s$.
\end{itemize}
A crucial consideration from the point of view of robustness of the results is to ensure that the obtained data pertain to the desired particle. In general, any lattice interpolating operator creates all states with the chosen quantum numbers, i.e., the desired hadron, but also its excitations. The latter need to be suppressed to isolate the ground state hadron. A spectral decomposition of the correlation functions implies that the excitations are exponentially suppressed and thus, the large-time two-point function receives contributions effectively only from the ground state. In the three-point function, excitations are suppressed the large distance between the insertion time, $\tau$, and the source/sink time. Hence, it is essential that the temporal separation $t_s$ is sufficiently large. To ensure the ground-state dominance, the ratio of Eq.~\eqref{eq:ratio} should be evaluated at several values of $t_s$ and the region $0\ll\tau\ll t_s$ be fitted, until convergence in $t_s$ is observed. These are the so-called plateau or one-state fits, in which all excitations need to be suppressed by large temporal distances. Alternatively, if three-point functions have been computed with several $t_s$ values, one can perform two-state (or, more generally, multi-state) fits and extrapolate to infinite $t_s$. 

\vspace*{1mm}\noindent\textbf{Enhancing signal quality}\\
We note that several techniques are typically used to enhance the signal quality of the underlying correlators.
The most crucial aspect of reliable GPDs extractions is to access as large nucleon boost in the direction of the Wilson line, $P_3$, as possible.
The obstacle to this is an exponentially worsening signal with increasing $P_3$. This problem can be alleviated by employing momentum smearing~\cite{Bali:2016lva} on quark fields, which modifies the standard Gaussian (Wuppertal) smearing~\cite{Alexandrou:1992ti} to enhance overlap with boosted ground states, as opposed to zero momentum ones. It is important to emphasize that momentum smearing does not solve the exponentially hard problem, but effectively slows down the exponential aggravation of the signal. In practice, most modern calculations employ hadron boosts of up to around 2 GeV. Calculations significantly above this boost may have compromised suppression of excited states by simulating with insufficiently large source-sink separations, with unknown systematic bias.
However, it is worth to note that a new method is being explored for accessing larger boosts, based on a kinematic enhancement of interpolating operators \cite{Zhang:2025hyo}, with promising results for PDFs \cite{Reitinger:2026hta}.
In addition to momentum smearing, other signal optimization strategies are used, with details depending on the collaboration, e.g., APE smearing~\cite{APE:1987ehd} on gauge fields and stout smearing~\cite{Morningstar:2003gk} on Wilson lines.

\vspace*{1mm}\noindent\textbf{Flavor combinations}\\
In principle, GPDs of any quark flavor can be accessed in LQCD. However, in general, flavor separation is difficult for two main reasons. Firstly, direct evaluations of MEs of individual quark flavors are computationally demanding due to the necessity to include quark-disconnected diagrams, which are significantly more noisy than quark-connected parts and require much larger statistics. Hence, most lattice studies are restricted to isovector combinations like $u-d$, in which disconnected loops cancel out in the isospin limit. The flavor-singlet combinations (e.g., the isoscalar combination $u+d$) or flavor-decomposed distributions, apart from requiring these disconnected diagrams, additionally mix with the gluon sector, significantly complicating the analysis and requiring dedicated tools. As such, they remain much less explored in lattice computations, with the complete flavor separation for GPDs still to be attempted.

\vspace*{1mm}\noindent\textbf{Symmetric and asymmetric frames of reference}\\
Finally, we comment on the frame of reference used for accessing MEs of Eq.~\eqref{eq:latME}. The standard method uses the symmetric frame, i.e., with a symmetric distribution of the momentum transfer between the source and the sink,
\begin{equation}
\mathbf{P}_i= \mathbf{P} - \frac{\mathbf{\Delta}}{2} = \bigg(\frac{-\Delta_1}{2}, \frac{-\Delta_2}{2},P_3 - \frac{\Delta_3}{2}\bigg), \qquad
\mathbf{P}_f= \mathbf{P} + \frac{\mathbf{\Delta}}{2} = \bigg(\frac{\Delta_1}{2}, \frac{\Delta_2}{2},P_3 + \frac{\Delta_3}{2}\bigg)\,.
\label{eq:sym}
\end{equation}
Accordingly with the above description of the fixed-sink sequential method of constructing the all-to-all propagator (and hence, the three-point function necessary to calculate the ME), the inevitable consequence of this setup is the need for separate calculations whenever the momentum transfer is changed. Obviously, this leads to prohibitively large computational cost when aiming for full mapping of the $t$ and $\xi$ dependence of GPDs.
To avoid this bottleneck, the natural idea is to keep the sink momentum fixed and assign the whole momentum transfer to the source,
\begin{equation}
\mathbf{P}_f  = \left(0,0,P_3\right),\quad
\mathbf{P}_i= \mathbf{P}_f - \mathbf{\Delta} = \left(-\Delta_1, -\Delta_2,P_3 - \Delta_3\right)\,,
\label{eq:asym}
\end{equation}
which defines an asymmetric frame of reference.
With a clear computational gain from this choice, the price to pay is the more complicated kinematics when relating asymmetric-frame MEs to GPDs, which we discuss next.

\vspace*{1mm}\noindent\textbf{From MEs to coordinate-space quasi-GPDs}\\
The matrix element related to the Dirac structure $\Gamma$ can be parametrized using the following trace:
\begin{equation}
\Pi^\Gamma_\kappa=K\mathrm{Tr}\bigg[\mathcal{P}_{\kappa}\bigg(\frac{-i\slashed{P}_f+M}{2M}\bigg)F_\Gamma^f\bigg(\frac{-i\slashed{P}_i+M}{2M}\bigg)\bigg], \qquad \kappa = 0,1,2,3,
\label{eq:trace}
\end{equation}
where  $K$ is a kinematic factor resulting from the normalization of a hadron state,
\begin{equation}
    K = \frac{2M^2}{\sqrt{E_fE_i(E_f+M)(E_i+M)}},
\end{equation} 
with $E_{i/f}$ being initial/final-state energies. In the symmetric frame, $E_i=E_f=E$ and $K$ simplifies to $2M^2/E(E+M)$.

For example, the standard unpolarized twist-2 GPDs are related to MEs with insertions $\gamma_0$ and the projectors $\mathcal{P}_0$ and $\mathcal{P}_{1/2}$, depending on the transverse components of momentum transfer,
\begin{equation}
H=h_1 \Pi^{\gamma_0}_0 + h_2 \Pi^{\gamma_0}_{1/2},\quad   
E=e_1 \Pi^{\gamma_0}_0 + e_2 \Pi^{\gamma_0}_{1/2},
\end{equation}
where the coefficients $h_i$, $e_i$ come from continuum parametrizations like Eq.~\eqref{eq:unpolGPD}. The coefficients are frame-dependent and more complicated in the asymmetric-frame case.
However, it is useful to parametrize $F_\Gamma^f$ using a theoretical tool of Lorentz-invariant (LI) amplitudes \cite{Bhattacharya:2022aob}. In the vector case, up to twist-4, the decomposition can be taken as\footnote{The decomposition is not unique, i.e., different Lorentz structures can be used, but the number of independent terms is fixed for a given channel. For example, Ref.~\cite{HadStruc:2024rix}, employs partially different Lorentz structures.}
\begin{align}
F^{\mu}(z,P,\Delta)  = \bar{u}(P_f,\lambda_f) \bigg [ \dfrac{P^{\mu}}{M} A_1 & + M z^{\mu} A_2 + \dfrac{\Delta^{\mu}}{M} A_3 + i M \sigma^{\mu z} A_4 \nonumber\\
&+ \dfrac{i\sigma^{\mu \Delta}}{M} A_5 
+ \dfrac{P^{\mu} i\sigma^{z \Delta}}{M} A_6
+ M z^{\mu} i\sigma^{z \Delta} A_7 + \dfrac{\Delta^{\mu} i\sigma^{z \Delta}}{M} A_8  \bigg ] u(P_i,\lambda_i),
\end{align}
where $u/\bar{u}(P,\lambda)$ denote Dirac spinors for momentum $P$ and polarization $\lambda$, $A_i$ ($i=1,\ldots,8$) are the LI amplitudes (their (suppressed above) dependence is on the Lorentz scalars $z\cdot P$, $z\cdot \Delta$, $z^2$ and $\Delta^2$) and $\sigma^{\mu \nu} \equiv \tfrac{i}{2} (\gamma^\mu \gamma^\nu - \gamma^\nu \gamma^\mu)$,  
$\sigma^{\mu z} \equiv \sigma^{\mu \rho} z_\rho$, 
$\sigma^{\mu \Delta} \equiv \sigma^{\mu \rho} \Delta_\rho$, $\sigma^{z \Delta} \equiv \sigma^{\rho \tau} z_\rho \Delta_\tau$.
The usefulness of the amplitude formalism is closely related to the practicals of lattice calculations.
With the fixed-sink sequential method, one can obtain all gamma structures in combination with all parity projectors at the same cost\footnote{Neglecting the comparatively low cost of contractions to yield MEs.}. Thus, the lattice data can fully determine all amplitudes in a single calculation. Then, the amplitudes can be related to all GPDs up to twist-4 and, moreover, one can construct alternative definitions of quasi-GPDs, with generally different convergence properties to light-cone GPDs (see below). Continuing the illustration for twist-2 unpolarized GPDs, the standard definition of GPDs leads to the following relations between $H$ and $E$ GPDs (at $\xi=0$, for simplicity) and LI amplitudes: \cite{Bhattacharya:2022aob}
\begin{equation}
\label{eq:HEsym}
H^s(A_i)   =  
A_1 + \frac{\Delta_\perp^2}{2 P_3}{zA_6},\qquad
E^s(A_i)   =  
- A_1  + 2 A_5 + \frac{\left( 4 E^2 - \Delta_\perp^2\right)}{2P_3}{zA_6},
\end{equation}
\begin{equation}
\label{eq:HEasym}
H^a(A_i) = 
A_1 + \frac{ (\Delta_\perp^2-(E_f-E_i)^2)}{2 P_3}{zA_6},\qquad
E^a(A_i) = 
- A_1 + 2 A_5 + \frac{\left(4E_f^2 - (E_f+E_i)(E_f-E_i) - \Delta_\perp^2 \right)}{2P_3}{zA_6},
\end{equation}
where $\Delta_\perp^2=\Delta_1^2 + \Delta^2_2$ and the upper index denotes the symmetric/asymmetric frame ($s/a$).
It is clear that the fact that $E_f\neq E_i$ in the asymmetric frame complicates the kinematic coefficients\footnote{The asymmetric-frame expressions in Ref.~\cite{Bhattacharya:2022aob} are written including the $A_3,A_4,A_8$ amplitudes. However, all these amplitudes vanish identically at $\xi=0$ due to symmetry properties.}.
The standard light-cone definition of GPDs leads to coefficients that only involve the plus and minus components of vectors like $P$, $\Delta$ and $z$, e.g., $P^+z^-$. Since the 4-vector $z$ has only the minus component non-zero in the definition of GPD correlators ($z^2=0$), the expressions for light-cone GPDs can be written in a Lorentz-invariant manner, e.g., $P^+z^-=P\cdot z$. For quasi-GPDs, one can generalize to non-zero $z^2$ and the Lorentz-invariant definition of quasi-GPDs reads:
\begin{equation}
\label{eq:HELI}
H(A_i) =  A_1 ,\qquad E(A_i) = - A_1 + 2 A_5 + 2 P_3 zA_6 .
\end{equation}
The Lorentz invariance of these quasi-GPDs implies that $H$ and $E$ no longer need a frame index. It is manifest that with respect to the standard definitions, the LI ones have the dependence on $zA_6$ removed ($H$) or reduced ($E$).
It is interesting to consider the MEs, $\Pi^\Gamma_\kappa$, that enter these definitions. As mentioned above, the standard definition utilizes $\Pi^{\gamma_0}_0$ and $\Pi^{\gamma_0}_{1/2}$. The LI one, even though it reduces the dependence on the amplitudes, additionally utilizes $\Pi^{\gamma_{1/2}}_3$ in both frames and also $\Pi^{\gamma_{1/2}}_0$ and $\Pi^{\gamma_{1/2}}_{2/1}$ in the asymmetric one. The gamma insertions $\gamma_{1/2}$ are related to twist-3 GPDs. Hence, it is clear that the LI definitions have modified HTEs and thus, an altered, possibly improved convergence to the light-cone GPDs.

The generalization to $\xi\neq0$ is straightforward, see Ref.~\cite{Chu:2025kew}, implying the LI definitions
\begin{equation}
\label{eq:HELIxi}
H(A_i) =  A_1-2\xi P_3 ,\qquad E(A_i) = - A_1 + 2\xi P_3 + 2 A_5 + 2 P_3 zA_6 - 4\xi P_3 zA_8,
\end{equation}
which, obviously, reduce to Eqs.~\eqref{eq:HELI} in the $\xi=0$ limit.
However, see Sec.~\ref{sec:asym} for a discussion of the special case of purely longitudinal momentum transfer.

The parametrization of the axial-vector channel, $\Gamma=\gamma^\mu\gamma_5$, can be taken as: \cite{Bhattacharya:2023jsc}
\begin{align}
\widetilde{F}^{\mu} (z, P, \Delta)
 = \bar{u}(P_f,\lambda_f) \bigg [ \dfrac{i \epsilon^{\mu P z \Delta}}{M} \widetilde{A}_1 &+ \gamma^{\mu} \gamma_5 \widetilde{A}_2 + \gamma_5 \bigg ( \dfrac{P^\mu}{M} \widetilde{A}_3 + M z^\mu \widetilde{A}_4 + \dfrac{\Delta^\mu}{M} \widetilde{A}_5 \bigg ) \nonumber \\[0.1cm]
& \hspace{1.65cm} + M \slashed{z}\gamma_5 \bigg ( \dfrac{P^\mu}{M} \widetilde{A}_6 + M z^\mu \widetilde{A}_7 + \dfrac{\Delta^\mu}{M} \widetilde{A}_8 \bigg )\bigg ] u(P_i, \lambda_i) \, ,
\label{eq:helicity}
\end{align}
with $\epsilon^{\mu P z \Delta}=\epsilon^{\mu \alpha \beta \gamma} P_\alpha z_\beta \Delta_\gamma$. The LI amplitudes in this case are denoted by $\widetilde{A}_i$ and similarly to the vector case, there are 8 of them, regardless of the chosen basis.
The standard definition of helicity GPDs implies the following relation with the amplitudes at zero skewness,
\begin{equation}
\label{eq:Htilde}
\widetilde{H}(\widetilde{A}_i) = \widetilde{A}_2 + z P_3 \widetilde{A}_6 - M^2 z^2 \widetilde{A}_7 \,,
\end{equation}
related to the ME $\Pi^{\gamma_5\gamma_3}_3$. The other helicity GPD, $\widetilde{E}$, cannot be accessed at $\xi=0$ due to a vanishing kinematic coefficient.
It is worth to mentioned that in this case, the standard definition is already LI. An alternative one (also LI), proposed in Ref.~\cite{Bhattacharya:2023jsc}, coincides with Eq.~\eqref{eq:Htilde}, except for having the amplitude $\widetilde{A}_7$ removed. Interestingly, even though the two definitions just differ by one term, concerning the higher-twist amplitude $\widetilde{A}_7$, the alternative definition of $\widetilde{H}$ uses MEs of $\Pi^{\gamma_5\gamma_{1/2}}_0$, $\Pi^{\gamma_5\gamma_{1/2}}_{1/2}$ and $\Pi^{\gamma_5\gamma_0}_3$, without any contribution of $\Pi^{\gamma_5\gamma_3}_3$.

Finally, the tensor sector needs 12 amplitudes $A_{Ti}$ and the parametrization can be written as: \cite{Bhattacharya:2025yba}
\begin{align}
F&^{[i\sigma^{\mu\nu}\gamma_5]} (z, P, \Delta)
 \!=\! \bar{u}(P_f,\lambda_f) \bigg [P^{[\mu}z^{\nu]}\gamma_5A_{T1} + \frac{P^{[\mu}\Delta^{\nu]}}{M^2}\gamma_5A_{T2} + z^{[\mu}\Delta^{\nu]}\gamma_5A_{T3} + \gamma^{[\mu}\bigg(\frac{P^{\nu]}}{M}A_{T4} + Mz^{\nu]}A_{T5} + \frac{\Delta^{\nu]}}{M}A_{T6}\bigg)\gamma_5 \nonumber \\[0.1cm]
&  + M\slashed{z}\gamma_5\bigg(P^{[\mu}z^{\nu]}A_{T7} + \frac{P^{[\mu}\Delta^{\nu]}}{M^2}A_{T8} + z^{[\mu}\Delta^{\nu]}A_{T9}\bigg)+i\sigma^{\mu\nu}\gamma_5A_{T10}  +i\epsilon^{\mu\nu Pz}A_{T11} + i\epsilon^{\mu\nu z\Delta}A_{T12} \bigg ] u(P_i,\lambda_i) \, ,
\label{eq:tensor}
\end{align}
where $A^{[\mu}B^{\nu]} = A^{\mu}B^{\nu}-A^{\nu}B^{\mu}$, $\epsilon^{\mu\nu Pz} = \epsilon^{\mu\nu\alpha\beta}P_{\alpha}z_{\beta}$, and $\epsilon^{\mu\nu z\Delta} = \epsilon^{\mu\nu\alpha\beta}z_{\alpha}\Delta_{\beta}$.
The standard definition of quasi-GPDs in the symmetric frame, at $\xi=0$, leads to:
\begin{align}
\label{eq:HEtensor}
    H_T^{s} = -2A_{T2}\bigg(1 + \frac{P^2}{M^2}\bigg) + A_{T4} + A_{T10}&,\qquad
    E_T^{s} = 2A_{T2} - A_{T4}, \nonumber\\
    \widetilde{H}_T^{s} = -A_{T2} - z A_{T12}\frac{M^2}{P_3}&,\qquad
    \widetilde{E}_T^{s} = -2A_{T6}, \,
\end{align}
employing only the twist-2 structures $\sigma_{3T}$, namely the MEs of $\Pi^{i\sigma_{31}\gamma_5}_2$, $\Pi^{i\sigma_{32}\gamma_5}_1$, $\Pi^{i\sigma_{31}\gamma_5}_0$, $\Pi^{i\sigma_{32}\gamma_5}_0$.
Interestingly, the alternative LI definition leads to a difference only in $\widetilde{H}_T$ (at $\xi\neq0$, also the other transversity GPDs differ),
\begin{align}
\label{eq:HEtensorLI}
H_T = -2A_{T2}\bigg( 1+ \frac{P^2}{M^2}\bigg) + A_{T4} + A_{T10}&,\qquad
E_T = 2A_{T2} -A_{T4}, \nonumber\\
\widetilde{H}_T = -A_{T2}&,\qquad
 \widetilde{E}_T = -2A_{T6},
\end{align}
where the cancelation of the $A_{T12}$ amplitude proceeds similarly as in the unpolarized case, i.e., by including the higher-twist MEs of $\Pi^{i\sigma_{01}\gamma_5}_2$ and/or $\Pi^{i\sigma_{02}\gamma_5}_1$.

\vspace*{1mm}\noindent\textbf{Renormalization}\\
Before proceeding with the extraction of physical information, the extracted coordinate-space GPDs need to be renormalized.
There are standard logarithmic divergences and, in addition, power divergences from the Wilson line \cite{Ji:2015jwa,Ishikawa:2017faj, Constantinou:2017sej}.
For quasi-distributions, the first non-perturbative program for renormalization of these divergences was based on a variant of the regularization-independent momentum subtraction (RI-MOM)~\cite{Martinelli:1994ty} tailored for non-local operators~\cite{Alexandrou:2017huk} for quasi-distributions.
Pseudo-distributions, in turn, were proposed to be renormalized in a ratio scheme~\cite{Orginos:2017kos}.
While the latter is still always employed for SDF, there have been additional proposals on the LaMET side, the hybrid scheme \cite{Ji:2020brr} or the related self-renormalization scheme \cite{LatticePartonLPC:2021gpi}.

In particular the former is becoming the standard choice in recent investigations.
The hybrid scheme combines non-perturbative renormalization based on the ratio scheme\footnote{Other non-perturbative schemes, such as RI/MOM, can also be used, but the ratio scheme is most commonly applied.} at small distances with an explicit extraction of the linear divergence and a physics-motivated extrapolation at larger distances.
However, a major theoretical challenge arises from the renormalon ambiguity \cite{Braun:2018brg,Liu:2020rqi,Holligan:2023rex,Zhang:2023bxs,Braun:2024snf}, which stems from the factorial growth of the perturbative expansion at large orders triggered by the linear divergence ($1/a$) in the self-energy of the Wilson line. This divergence introduces an IR pole in the Borel transform of the subtraction kernel, creating an inherent $\mathcal{O}(\Lambda_{\rm QCD})$ ambiguity when defining the self-energy subtraction.
To eliminate this ambiguity, leading-renormalon resummation (LRR) \cite{Holligan:2023rex,Zhang:2023bxs} can be employed to identify and sum the dominant bubble-chain diagrams to all orders in perturbation theory. This allows for a principal-value prescription that systematically absorbs the unphysical linear divergence into the non-perturbative definition of the lattice matrix element.
Simultaneously, renormalization group resummation (RGR) \cite{Su:2022fiu} can be utilized to sum the large logarithmic corrections ($\ln z^2\mu^2$) that may become large at distances away from the inverse of the renormalization scale.
The combined application of both LRR and RGR leads to the most robust renormalized quasi-distributions, which can be subjected to the next stages of the LaMET procedure.

This concludes our discussion of coordinate-space GPDs extracted from the lattice. These objects are defined in Euclidean spacetime and at this stage, are not yet related to physical, light-cone quantities. Their translation to the latter is the topic of the next subsection, together with the (optional) transformation to momentum space.

\subsection{From Euclidean coordinate-space GPDs to light-cone momentum-space GPDs or their moments}
\subsubsection{Large momentum effective theory and quasi-distributions}
\vspace*{1mm}\noindent\textbf{Quasi-distributions}\\
Historically, the first approach that utilized the non-local Euclidean matrix elements defined above to extract light-cone partonic distributions was LaMET, proposed by Ji \cite{Ji:2013dva,Ji:2014gla}. The first stage of LaMET is to reconstruct the $x$ dependence, yet in Euclidean metric.
This defines quasi-distributions, generically denoted here as $\widetilde{G}^f(x,P_3)$ for flavor $f$,
\begin{align}
\label{eq:Fourier_quasi}
G^f(z,P_3) =\int_{-1}^{1}dx\,e^{-ixzP_3}\,\widetilde{G}^f(x,P_3)\,,
\end{align}
where $G^f(z,P_3)$ is the underlying coordinate-space renormalized GPD and we suppress all arguments except for the ones indicating the coordinate/momentum space and the hadron boost in the Wilson line direction. 
The above equation can be decomposed into the real and imaginary parts, with the former being related solely to the valence distribution (cosine part) and the latter to the combination of valence and twice the sea (sine part).

\vspace*{1mm}\noindent\textbf{Inverse problem and its regularization}\\
While Eq.~\eqref{eq:Fourier_quasi} expresses a well-defined mathematical relation between coordinate-space renormalized GPDs and their momentum-space, $x$-dependent counterparts, this Fourier transform step is very delicate.
Namely, the left-hand-side are discrete, truncated and statistically noisy lattice data, for some integer values of $z/a$ until $\zmax/a$.
The right-hand-side, in turn, is a continuous distribution and hence, the problem of its extraction is an ill-posed inverse problem.
Solving for $\widetilde{G}^f(x,P_3)$ means having to ``invert'' the integral of Eq.~\eqref{eq:Fourier_quasi}. 
Obviously, such a solution is non-unique and needs an introduction of an additional criterion.
In the context of partonic distributions, the problem was first discussed in Ref.~\cite{Karpie:2019eiq} and some solutions to it have been offered and analyzed.
The possible methods include:
\begin{itemize}
\item Discrete Fourier transform -- Eq.~\eqref{eq:Fourier_quasi} can be discretized and a $\zmax$ cut can be adopted -- while the criterion is very simple, it typically leads to unphysical oscillatory behavior.
\item Backus-Gilbert (BG) method \cite{BackusGilbert} -- regulates the ill-posed inversion with a model-independent criterion: minimizing the statistical variance of the solution while maximizing spatial resolution; the obtained distribution is unique and the oscillatory behavior is significantly mildened, but the agreement between the original coordinate-space data and the inverse Fourier transform of the BG-reconstructed distribution is not guaranteed.
\item Bayesian approaches -- using physics-motivated prior knowledge to determine the most probable continuous distribution. For example, a functional form of the $x$-dependent distribution can be used a prior and serve as a default model. In addition, a regulator can be used that dictates how much deviations from the default model are penalized (e.g., a maximum entropy method or ``Bayesian reconstruction'') \cite{Karpie:2019eiq}.
Another possibility is the Gaussian Process Regression (GPR) \cite{Alexandrou:2020tqq}, which assumes that the coordinate-space distribution (renormalized MEs) follow a Gaussian process prior, which pertains to its shape, smoothness and asymptotic behavior. Thus-regularized coordinate-space distribution can be subjected to an analytically-computable FT.
\item Neural networks -- used to parameterize continuous distributions with flexible, model-independent, non-parametric functional forms. The data-driven reconstruction minimizes human bias and ensures realistic uncertainties in regimes not well constrained by data. Apart from addressing the inverse problem, they can be used to serve additional purposes (see below).
\item Asymptotic extrapolation -- originally proposed in Ref.~\cite{Ji:2020brr} and developed in Ref.~\cite{Ji:2026vir}, this approach uses the physical property that spatial correlations must decay exponentially at sufficiently large distances; corrections to this behavior may be calculated employing heavy-quark effective theory, dispersive analysis, Lorentz symmetry, and heavy-flavor spectra. Fitting the data in the sub-asymptotic regime allows for physically robust extrapolation to infinite distances, which enables the use of a continuous FT.
\end{itemize}

\vspace*{1mm}\noindent\textbf{Matching to light-cone distributions}\\
The quasi-distribution, $\widetilde{G}^f(x,t,\xi,\mu,P_3)$, depends on the longitudinal momentum fraction $x$, the kinematic variables describing the momentum transfer, $t$ and $\xi$, the renormalization scale $\mu$ and the hadron boost $P_3$. The final required step is to factorize it into a light-cone distribution $G^f(x,t,\xi,\mu)$, a function that, in principle, no longer depends on the hadron boost:
\begin{align}
\label{eq:factor_quasi}
\widetilde{G}^f(x,t,\xi,\mu,P_3)=\int_{-1}^{1}\frac{dy}{|y|}C\left(\frac{x}{y},\frac{\xi}{y},\frac{\mu}{yP_3}\right)G^f(y,t,\xi,\mu)+\mathcal{O}\left(\frac{\Lambda^2_{\rm{QCD}}}{|x\pm\xi|^2P_3^2},\frac{\Lambda^2_{\rm{QCD}}}{|x\pm1|^2P_3^2},\frac{-t}{P_3^2}\right),
\end{align}
where $C(\ldots)$ is the perturbatively calculable momentum-space matching kernel.
The above factorization formula is valid up to power-suppressed $\mathcal{O}(1/P_3^2)$ corrections. The latter are enhanced close to $x=\pm\xi$ and $x=\pm1$, as well as when the invariant momentum transfer $t$ becomes comparable to $P_3^2$.
Thus, only some regions of the $x$ dependence can be reliably obtained via LaMET, namely ones that are at least $\Delta x\sim\mathcal{O}(\Lambda_{\rm{QCD}}/{P^z})$ away from the points $x=-1,-|\xi|,|\xi|,1$. With typical boosts achieved in contemporary lattice simulations, $\Delta x\sim0.15-0.2$. Assuming $\xi=1/4$ and $\Delta x=0.15$ for concreteness, this implies reliable LaMET results for $|x|\in[0,0.1]\cup[0.4,0.85]$, i.e., the central part of the ERBL region and the intermediate-$x$ DGLAP region. At zero skewness, the corresponding reliable region is $|x|\in[0.15,0.85]$. It is important to emphasize that this reliable region can be systematically enhanced by employing larger hadron boosts. However, the computational cost of accessing such larger boosts increases exponentially.
Alternative ways of augmenting LaMET in its unreliable regions are discussed below.

Historically, matching was performed on quasi-distributions renormalized in intermediate non-perturbative schemes, such as the RI/MOM scheme mentioned above. 
Thus, apart from matching to the light-cone, an additional task was to convert to the desired continuum scheme, i.e., the $\MSb$ scheme at a chosen renormalization scale.
For GPDs, the relevant framework was developed in Ref.~\cite{Liu:2019urm}, which was heavily used until the advent of the hybrid scheme.

The hybrid scheme \cite{Ji:2020brr} was introduced in order to overcome the deficiencies of the RI/MOM scheme, such as the presence of residual divergences at large distances \cite{Zhang:2020rsx}.
The hybrid matching for GPDs was developed in Ref.~\cite{Yao:2022vtp}.
However, similarly to the renormalization phase, the matching procedure within the hybrid framework is also plagued by the renormalon ambiguity, originating from the  perturbative matching kernel and generating $\mathcal{O}(\LambdaQCD/xP_3)$ power corrections. 
The leading-renormalon resummation framework developed in Refs.~\cite{Holligan:2023rex,Zhang:2023bxs} and mentioned above for renormalization includes also ingredients necessary at the matching stage.
Likewise, the renormalization-group resummation work of Ref.~\cite{Su:2022fiu} discusses the required modifications in the hybrid matching.
Finally, the matching kernel may also introduce large logarithms near the kinematic threshold ($x\rightarrow1$), which arise from an interplay of collinear and soft divergences and can also be resummed \cite{Gao:2021hxl,Ji:2024hit,Holligan:2025baj}.
Overall, the simultaneous application of LRR, RGR, and threshold resummation ensures that the hybrid matching framework yields a theoretically most robust reconstruction of the GPDs across the domain of validity of LaMET.

\subsubsection{Short distance factorization and pseudo-distributions}
\vspace*{1mm}\noindent\textbf{Pseudo-distributions}\\
An alternative procedure that can be applied to the same lattice data is the pseudo-distribution approach. The most significant difference is the reversed order of momentum space reconstruction and factorization from Euclidean to light-cone distributions, the latter coming as the first step, yet in coordinate space, at short distances (SDF).
Moreover, the $x$ dependence reconstruction is optional, as SDF gives clean access to moments of GPDs.
Such an access is impossible within LaMET, due to the presence of unreliable regions in $x$.

The pseudo-distribution approach typically operates on ratio-renormalized MEs, where the ratio is constructed to cancel all divergences between the numerator and the denominator. Since these divergences do not depend on momentum transfer, a GPD bare matrix element, $G^f(z,\nu)$, can be divided by a zero-momentum PDF at the same Wilson line length. This gives rise to a pseudo-GPD, also referred to as a pseudo-ITD (Ioffe time distribution), with Lorentz-invariant arguments $z^2$ and the Ioffe time \cite{Ioffe:1969kf}, $\nu\equiv P\cdot z$. Upon the standard choice of the 4-vector $z$ taken along the third direction, we write the pseudo-GPD as $\mathcal{G}^f(z,\nu,t,\xi)$.

\vspace*{1mm}\noindent\textbf{Matching to light-cone ITDs}\\
The pseudo-GPD is factorized according to: 
\begin{align}
\mathcal{G}^f(z,\nu,t,\xi) =\int_{-1}^{1}du\, C_\nu\left(u\nu,z\mu\right) \, G^f(u\nu,t,\xi,\mu)+\mathcal{O}\left(z^2\Lambda^2_{\rm{QCD}},z^2t\right).
\end{align}
The matching kernel is written with a subscript indicating that it acts in coordinate (Ioffe time) space.
$C_\nu$ is related to the corresponding momentum-space matching kernel by a convoluted integral transform that physically convert between the short-distance coordinate resolution scale ($z$) and a large-momentum scale ($P_3$).
The above factorization relation is, analogously to the LaMET case, valid up to power corrections, which relate the short-distance scale $z$ with the non-perturbative scale $\Lambda_{\rm QCD}$ and the scale given by the momentum transfer, $t$.
The power corrections are under good control only when $z$ is small, i.e., $\zmax\lesssim0.2-0.3$ fm. In other words, SDF provides reliable information only for a rather small Ioffe time segment, ranging from 0 to $\nu_{\rm{max}}=\zmax P_3^{\rm max}$,
where $P_3^{\rm max}$ is the maximum attained hadron boost.

\vspace*{1mm}\noindent\textbf{Reconstruction of $x$ dependence}\\
This restriction makes it difficult to realize the last step of the procedure of reconstructing an $x$-dependent GPD.
While formally given by the FT (again suppressing arguments other than the ones relevant for this step),
\begin{align}
\label{eq:Fourier_pseudo}
G^f(\nu) =\int_{-1}^{1}dx\,e^{-ix\nu}\,G^f(x)\,,
\end{align}
the restriction to $\nu\leq\nu_{\rm max}$ makes the inverse problem significantly more ill-defined, unless $P_3^{\rm max}$ is very large.
However, the light-cone ITDs provided by the factorization step are already physical objects of interest, which characterize the hadron's structure in Ioffe time space, with $\nu$ being Fourier-conjugate to $x$.

\vspace*{1mm}\noindent\textbf{Extraction of Mellin moments}\\
Alternatively, this SDF-derived information can be expressed as Mellin moments of GPDs.
The reduced-ITD, corresponding to a GPD $G^f$, can be written as a sum over its moments:
\begin{equation}
\label{eq:sum_moments}
    \mathcal{G}^f(z,\nu,t,\xi)=\sum_{n=0}^{\infty}\frac{(-i\nu)^n}{n!}\,\mathcal{G}_{n+1}^f(z,t,\xi)\,,
\end{equation}
where
\begin{equation}
    \mathcal{G}_n^f(z,t,\xi)=\int_{-1}^{1}dx\,x^{n-1} \mathcal{G}^f(x,t,\xi)
\end{equation}
are Mellin moments of the considered pseudo-GPD, which require matching to light-cone moments.
Analogously, one can define moments of the light-cone distributions, $G_n(t,\xi)$.
Below, we concentrate on the nucleon\footnote{The picture is simpler for light pseudoscalar mesons, such as the pion, for which significant amount of lattice work has recently been done, see below.} and its unpolarized moments.
The light-cone moments of the $H$ and $E$ GPDs can be written in terms of polynomiality relations involving GFFs, $A_{n,m}(t)$ and $B_{n,m}(t)$, and moments of the D-term, $C_n(t)$,
\begin{align}
\int_{-1}^{1} dx \, x^n H(x, t, \xi) = \sum_{\substack{k=0 \\ \text{even}}}^{n} A_{n+1, k}(t) \, \xi^k + \bmod(n, 2) \cdot C_{n+1}(t) \, \xi^{n+1},
\end{align}
\begin{align}
\int_{-1}^{1} dx \, x^n E(x, t, \xi) = \sum_{\substack{k=0 \\ \text{even}}}^{n} B_{n+1, k}(t) \, \xi^k - \bmod(n, 2) \cdot C_{n+1}(t) \, \xi^{n+1},
\end{align}
where $\bmod(n, 2)$ is zero/one for even/odd $n$.
Thus, the first moments of $H/E$ are just the GFFs $A_{1,0}(t)$ and $B_{1,0}(t)$, while higher moments are related to $A/B_{n,0}(t)$ and the skewness-dependent $A/B_{n,m}(t)$ ($0<m<n$, $m$ even) and moments of the D-term, $C_n$ (only even $n$), with opposite signs for $H/E$.

The matching relation between pseudo- ($\mathcal{G}_n$) and light-cone ($G_n$) moments can be schematically written as
\begin{align}
\mathcal{G}_n(z,t,\xi)=
\sum_{\substack{k=0 \\ \text{even}}}^{n} \mathcal{C}_{nk}(\xi,\alpha_s(\mu),z\mu)\,
G_k(\mu,t,\xi)
+\mathcal O\left(
z^2\Lambda_{\rm QCD}^2,
z^2t
\right),    
\end{align}
where $\mathcal{C}_{nk}$ are perturbative matching coefficients (Wilson coefficients). The $\mathcal{C}$ matrix, related to the $C_\nu$ ITD matching kernel above, has skewness-independent diagonal entries and the $\xi$ dependence enters off the diagonal (leading to the mixing of orders of pseudo- and light-cone moments). The Wilson coefficients can also be RGR-improved, see Ref.~\cite{HadStruc:2024rix} for more details. Similarly to the matching of ITDs, this matching relation is valid up to HTEs of $\mathcal O\left(z^2\Lambda_{\rm QCD}^2,
z^2t\right)$ and effectively, accessing higher moments is possible when large-$\nu$ ITDs can be reliably included at small $z$, i.e., if the hadron boost is sufficiently large.

\subsubsection{LaMET vs.\ SDF -- summary}
\begin{figure}[t!]
    \centering
    \includegraphics[scale=1]{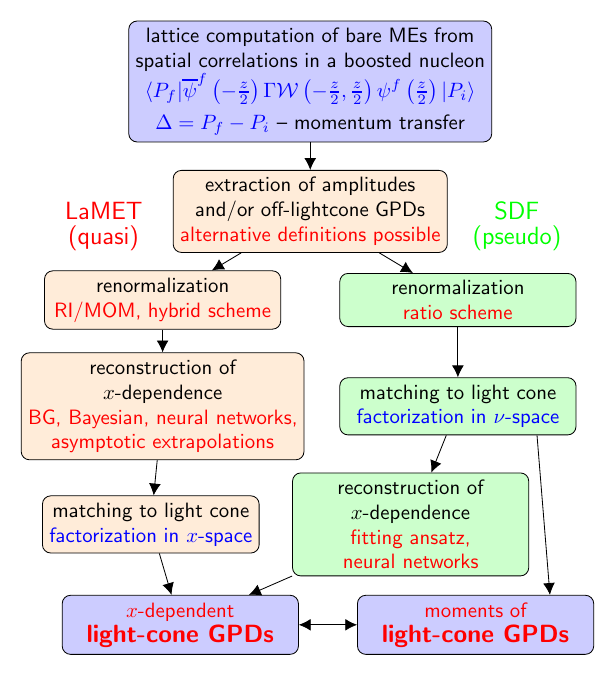} 
    \caption{Schematic illustration of the steps needed to go from lattice-calculated bare MEs to final light-cone GPDs or their moments.}    \label{fig:quasi_pseudo}
\end{figure}

In principle, the information extracted from SDF should be consistent with the one from LaMET.
This statement becomes obvious in the infinite momentum frame, where the two approaches are strictly equivalent \cite{Izubuchi:2018srq}.
However, at a finite hadron boost, LaMET and SDF are affected by different systematic effects and thus, have different ranges of applicability.
We summarize the two approaches with a schematic illustration of the steps involved in the LaMET and SDF extractions of light-cone GPDs or their moments, see Fig.~\ref{fig:quasi_pseudo}.
The chart emphasizes that the lattice input is the same in both approaches and the main difference is the factorization being performed either in momentum or coordinate space. While both approaches can, in principle, reconstruct the $x$ dependence, the factorization in momentum space makes it more natural within LaMET, although in a restricted $x$ range.
In turn, the factorization in coordinate space (SDF) makes the $x$ dependence reconstruction model-dependent, with GPD Mellin moments the natural objects for a reliable extraction.
Overall, it needs to be remembered that the maximum hadron boost is the crucial limitation in both frameworks, translating to the reliable $x$ regime in LaMET and to the reliable highest order moment in SDF.

\section{Review of results}
\label{sec:results}
\subsection{Exploratory works in the symmetric frame}
The initial explorations of GPDs from LaMET were performed in the symmetric (Breit) frame of reference.

In 2019, Chen et al.~\cite{Chen:2019lcm} presented the first lattice calculation\footnote{Lattice setup: clover valence fermions on an $N_f=2+1+1$ staggered (HISQ) sea, $a\approx0.12$ fm, $m_\pi\approx310$ MeV, $P_3^{\rm max}\approx1.7$ GeV.} of pion valence $H$ GPD with two nonzero values of momentum transfer at zero skewness.
Results were renormalized in the RI/MOM scheme and matched to the light cone. The study proved the feasibility of pion GPDs determinations from the lattice and reproduced expected qualitative features, such as the suppression of GPDs with increasing $-t$.

Still in 2019, preliminary analyses were reported also for the nucleon \cite{Alexandrou:2019dax}, by the Extended Twisted Mass Collaboration (ETMC).
The full computation was accomplished the next year \cite{Alexandrou:2020zbe}\footnote{Lattice setup: $N_f=2+1+1$ twisted mass fermions at maximal twist,  $a\approx0.093$ fm, $m_\pi\approx260$ MeV, $P_3^{\rm max}\approx1.7$ GeV.}, presenting results
for two sectors -- unpolarized GPDs ($H$ and $E$) and helicity GPDs ($\tilde H$ and $\tilde E$).
Two values of momentum transfer were used, $-t=0.69$ GeV$^2$ at zero skewness\footnote{In this case, $\tilde E$ is inaccessible due to a vanishing kinematic coefficient.} and $-t=1.02$ GeV$^2$ at $\xi=1/3$. Thus, this study offered the first glimpse into the ERBL region of GPDs, also revealing the significant difficulties for the neighborhood of $x=\xi$ -- the GPDs can have a discontinuity of their first derivative, but should be continuous at $x=\xi$. The observed jumps at $x=\pm\xi$ reflect the enhanced power corrections close to the boundaries between the ERBL and DGLAP regions.

A follow-up study in the same lattice setup appeared in 2021 \cite{Alexandrou:2021bbo}, addressing, for the first time, chiral-odd transversity GPDs. There are four leading-twist GPDs, of which $H_T$ was shown to have the best signal, while the results for $E_T$ and $\tilde{H}_T$ were found to be very noisy. Finally, $\tilde{E}_T$ vanishes at $\xi=0$, but was found to be consistent with zero also at $\xi=1/3$, $-t=1.02$ GeV$^2$. A clear conclusion of the paper was that the calculations of chiral-odd GPDs are challenging and require large statistics, apart from the leading $H_T$ GPDs, whose forward limit is the transversity PDF.

Independent exploratory studies of nucleon GPDs from a lattice calculation and LaMET matching were performed around the same time by Lin \cite{Lin:2020rxa,Lin:2021brq}\footnote{Lattice setup: clover valence fermions on an $N_f=2+1+1$ staggered (HISQ) sea, $a\approx0.09$ fm, $m_\pi\approx135$ MeV, $P_3^{\rm max}\approx2.2$ GeV.}, marking the first attempt at physical point extractions of GPDs. The two papers considered zero-skewness unpolarized ($H$,$E$) and helicity ($\tilde H$) GPDs at $-t=0.19,\,0.39,\,0.77,\,0.97$ GeV$^2$ and modeled the $t$-dependence of GPDs to take the Fourier transform to impact-parameter space and arrive at first tomographic pictures of the nucleon.
Another physical-point extraction of GPDs, for the pion, was reported by Lin in 2023 \cite{Lin:2023gxz},\footnote{Lattice setup: clover valence fermions on an $N_f=2+1+1$ staggered (HISQ) sea, $a\approx0.09$ fm, $m_\pi\approx135$ MeV, $P_3^{\rm max}\approx1.7$ GeV.} with the same $-t$ values as for the nucleon.  For the first time in the context of GPDs, this paper applied the hybrid renormalization, asymptotic extrapolation of renormalized MEs to infinite Ioffe time to allow for a well-defined discrete Fourier transform, and hybrid matching. Similarly to the earlier nucleon papers, a tomographic picture of the pion was obtained upon a Fourier transform to impact parameter space.

Later, a systematic study of renormalization and matching for nucleon GPDs was performed in Ref.~\cite{Holligan:2023jqh}, using the lattice data of Ref.~\cite{Lin:2020rxa} with the same range of $-t$ values at $\xi=0$ and additionally, $-t=0.23$ GeV$^2$ at $\xi=0.1$.
Renormalization was performed in the hybrid scheme, based on the ratio of boosted MEs and zero-momentum MEs to cancel divergences at short distances and an explicit subtraction of self-energy divergences in the Wilson line for long distances, aided by asymptotic extrapolations to infinite Ioffe time. Perturbative matching was performed to translate from hybrid-renormalized quasi-GPDs to light-cone GPDs. Importantly, at $\xi=0$, both renormalization and matching included LRR to remove the renormalon ambiguity and RGR of small-$x$ logarithmic terms. The whole procedure was performed at the 1-loop and the 2-loop level, allowing for assessments of possible pertubative truncation effects.
For $\xi\neq0$, 1-loop matching with only LRR could be applied, due to unavailable RGR and 2-loop expressions for this case.

In 2023, yet another symmetric-frame exploratory work appeared \cite{Bhattacharya:2023nmv} in the same lattice setup as the above ETMC works, targeting the elusive twist-3 GPDs at zero skewness. The authors considered the axial-vector channel, which includes four twist-3 GPDs, $\tilde{G}_1,\,\tilde{G}_2,\,\tilde{G}_3,\,\tilde{G}_4$, with $\tilde{G}_3$ vanishing at $\xi=0$. The first two of these GPDs are extracted as combinations with twist-2 helicity GPDs, $\tilde H+\tilde{G}_2$ and $\tilde E+\tilde{G}_1$, where the former could be disentangled by using previous results on $\tilde{H}$. The study found a very good signal of $\tilde H+\tilde{G}_2$ and $\tilde E+\tilde{G}_1$ and suppressed values of $\tilde{G}_4$. Even though the extractions are clearly contaminated by unaccounted for systematic effects, the results were shown to pass several consistency checks, such as the vanishing of the norm of all extracted twist-3 GPDs.

One can draw a few common conclusions from the above exploratory studies, which shaped the progress in next years:
\begin{itemize}
\item The feasibility of lattice extractions of GPDs was fully established in all twist-2 channels as well as one of the twist-3 sectors.
\item The computational bottleneck for full mapping of the kinematic domain of GPDs was the necessity to perform separate simulations at each value of momentum transfer.
\item The need to investigate various sources of systematic effects was established, concerning lattice-specific aspects (discretization effects, excited states, etc.) as well as other effects entering different steps of the extraction (renormalization, matching, $x$ dependence reconstruction), with desired improvements.
\end{itemize}

In the next subsection, we discuss the breakthroughs that came after these exploratory works, particularly related to the introduction of asymmetric frames of reference.

\subsection{Breakthrough of asymmetric frames and other developments}
\label{sec:asym}
The key development that changed the perspective of lattice extractions of GPDs was the introduction of asymmetric frames, allowing for simultaneous calculations of several momentum transfer values. In this subsection, we discuss this progress, together with other advancements of the field of lattice GPDs. We start with developments introduced in the context of LaMET and discuss separately the progress in SDF.

\subsubsection{LaMET}
Exploratory works in the symmetric frame naturally led to the question how to avoid the necessity of repeating all numerically intensive steps whenever a different momentum transfer vector is considered.
This triggered the work of Ref.~\cite{Bhattacharya:2022aob}\footnote{Lattice setup: $N_f=2+1+1$ twisted mass fermions at maximal twist,  $a\approx0.093$ fm, $m_\pi\approx260$ MeV, single $P_3\approx1.25$ GeV.}, leading to the formalism detailed in Section 2, in the unpolarized case. The purpose of the lattice study was to validate the framework, in particular to verify the frame independence of the extracted Lorentz-invariant amplitudes, $A_{1,\ldots,8}$. This was accomplished by an explicit comparison of these amplitudes between the symmetric frame and the asymmetric frame, for the same momentum transfer vector $(2,0,0)$ (plus permutations and opposite values of $P_3$). The kinematics of symmetric/asymmetric frames implies a slightly different value of $-t$ for both cases, 0.69 and 0.64 GeV$^2$, respectively, but the implied difference at the level of amplitudes or final GPDs was argued to be well below statistical precision. Moreover, the study showed that only $A_1$ and $A_5$ have large values, while all other amplitudes are subleading or compatible with zero. Finally, the paper compared the standard and LI definitions of GPDs, confirming the expectation that the former/latter are frame-dependent/independent.
Results complementing this work were shown in Ref.~\cite{Cichy:2023dgk}, which considered a wide range of $-t$ values, ranging from 0.17 to 2.24 GeV$^2$, all obtained from a single calculation. Importantly, the good level of statistical precision still holds well above $-t=1$ GeV$^2$, with only the largest $-t$ values showing deteriorated signal. However, the most important demonstration of this follow-up study concerned the convergence of standard/LI definitions to light-cone GPDs. Based on three values of $P_3=0.83,\,1.25,\,1.67$ GeV, it was shown that the convergence of the $H$ GPD is very similar in both definitions, but the LI $E$ GPD converges much faster than the standard $E$ GPD. Moreover, the latter evinces much better statistical precision. At the level of amplitudes, both $E$ definitions differ by the coefficient of the $A_6$ amplitude and thus, the better convergence results from suppressed power corrections present in $A_6$. At the level of matrix elements, this improvement corresponds to subtraction of correlated noise in the matrix elements of $\gamma_0$, $\gamma_1$ and $\gamma_2$ insertions.

The success of the asymmetric-frame formalism in the unpolarized sector triggered investigations into the polarized cases -- helicity \cite{Bhattacharya:2023jsc} and transversity \cite{Bhattacharya:2025yba}. Both of these papers developed the theoretical formalism, based on 8 amplitudes $\tilde{A}_i$ and 12 amplitudes $A_{Ti}$, respectively, and considered two alternative definitions of twist-2 GPDs. In the lattice part, realized in the same setup as Ref.~\cite{Bhattacharya:2022aob}, the axial-vector/tensor formalisms were validated by explicit comparisons of amplitudes and GPDs extracted from the symmetric/asymmetric frame and results were presented for a wide range of $-t$ from a single calculation.

The final extension so far, utilizing the same lattice setup in the numerical part, is the case of nonzero skewness for unpolarized GPDs \cite{Chu:2025kew}.
While the formalism of Ref.~\cite{Bhattacharya:2022aob} is general, it requires a special consideration of the kinematic setup of purely longitudinal momentum transfer. In such a setup, the number of independent amplitudes reduces from 8 to 3, with the three amplitudes becoming linear combinations of the eight general-setup amplitudes. As a consequence, it is no longer possible to disentangle the $H$ and $E$ GPDs. Instead, this setup gives access to their skewness-dependent combination, denoted $H^L$. Nevertheless, at typical lattice kinematics and moderate values of skewness, the combination is dominated by $H$, with 1-3\% admixture of $E$. Hence, the contamination is hidden in the typical statistical precision and $H^L$ can be considered to be equivalent to $H$ for practical purposes. Importantly, this longitudinal setup provides then access to very low values of $-t$, unattainable when transverse momentum transfer is present.
The lattice part accessed three values of the skewness, 1/7, 1/5 and 1/2 (of both positive and negative signs), at $-t$ ranging from 0.09 to 2.25 GeV$^2$. It verified the largest magnitudes of the amplitudes $A_1$ and $A_5$ and found suppressed, but possibly nonzero values for $A_2,\,A_6$ and $A_7$. It also demonstrated the influence of skewness at fixed or approximately fixed $-t$ and verified the symmetry between $\pm\xi$. Finally, reconstructions of $x$-dependent quasi-GPDs were performed and the latter were matched to the light cone. This provided further evidence of enhanced power corrections at $|x|\approx|\xi|$. This points to an important conclusion that the LaMET extraction should be complemented with additional input, which we discuss below.

The asymmetric-frame formalism was also put to use for pion GPDs at zero skewness \cite{Ding:2024saz}\footnote{Lattice setup: clover valence fermions on an $N_f=2+1$ staggered (HISQ) sea, $a\approx0.04$ fm, $m_\pi\approx160$ MeV (sea), $m_\pi\approx300$ MeV (valence), $P_3^{\rm max}\approx1.9$ GeV.}, addressing a broad range of $-t$ up to around 1.7 GeV$^2$ and arriving at impact-parameter space tomographic pictures.
An important ingredient of this paper with respect to the nucleon studies was the improvement of the renormalization, $x$ dependence reconstruction and matching setup.
Similarly to Ref.~\cite{Holligan:2023jqh}, discussed above (symmetric-frame lattice data),
the authors used hybrid renormalization, asymptotic extrapolations and hybrid matching with LRR and RGR improvements.

\subsubsection{SDF}
The alternative way to extract GPD-related physics from the lattice is to use factorization in coordinate space. We remind the Reader that this approach is based on the same non-local matrix elements as LaMET, but after renormalization, it skips the step of $x$ dependence reconstruction. Instead, ratio-renormalized MEs (pseudo-ITDs) are subjected to a matching procedure or expanded in Ioffe time. The former leads to matched ITDs, physical objects that can be further used to access light-cone GPDs upon a Fourier transform. The other way, expansion in $\nu$,  gives Mellin moments of GPDs from applying perturbative Wilson coefficients.

The first study that applied SDF to asymmetric-frame lattice data was Ref.~\cite{Bhattacharya:2023ays}\footnote{Lattice setup: $N_f=2+1+1$ twisted mass fermions at maximal twist,  $a\approx0.093$ fm, $m_\pi\approx260$ MeV, $P_3^{\rm max}\approx1.7$ GeV.}. It utilized the same lattice data as Ref.~\cite{Bhattacharya:2022aob} (including the symmetric frame) and extracted skewness-independent moments of isovector and isoscalar\footnote{Neglecting quark-disconnected diagrams and mixing with the gluon, both argued to be negligible at this level of precision.} $H$ and $E$ GPDs, using the standard and LI definitions.
Perturbative matching was done at the NNLO level, showing very small 2-loop effects, and adopting RGR improvement.
Results up to $A/B_{5,0}$ were obtained, with the fifth order moments on the verge of statistical significance.
The $t$ dependence was parametrized with a dipole fit and the $z$-expansion, both evincing good description of the obtained data.
Extrapolation to the forward limit yielded a determination of quark charges, momentum fractions, and the angular momentum contributions
to the proton spin.
Finally, the paper attempted a Fourier transform to impact parameter space, presenting tomographic pictures of the first four moments, for unpolarized quarks in an unpolarized, as well as in a transversely polarized nucleon.

After extension of the asymmetric-frame formalism to the axial-vector case, the same lattice setup was used to access Mellin moments of the $\tilde H$ GPD\footnote{We remind the Reader that $\tilde E$ has a vanishing kinematic coefficient at $\xi=0$.} \cite{Bhattacharya:2024wtg}. The matching setup was also analogous, but restricted to NLO due to the unavailability of NNLO expressions. Similarly to the unpolarized case, moments up to the fifth one were extracted. The results provided insights into the spin structure of the nucleon, including quark helicity, quark orbital angular momentum and spin-orbit correlation. The densities of these three observables were also depicted in impact parameter space.

The first extraction of unpolarized Mellin moments including their skewness dependence was performed by the HadStruc collaboration \cite{HadStruc:2024rix}\footnote{Lattice setup: $N_f=2+1$ clover fermions,  $a\approx0.094$ fm, $m_\pi\approx358$ MeV, $P_3^{\rm max}\approx1.4$ GeV.}.
They accessed moments up to the fourth one, including GFFs that contribute only at $\xi\neq0$, i.e., $A/B_{3,2}$, $A/B_{4,2}$, and $C_2$ (related to the $D$-term).
Perturbative matching was performed at NLO, in different variants of DGLAP leading-logarithm resummation.
The $t$ dependence was parametrized with a dipole fit and the $z$-expansion, leading to tomographic pictures of the extracted moments.

A similar study was performed also for the pion \cite{Gao:2025inf}\footnote{Lattice setup: clover valence fermions on an $N_f=2+1$ staggered (HISQ) sea, $a\approx0.04$ fm, $m_\pi\approx160$ MeV (sea), $m_\pi\approx300$ MeV (valence), $P_3^{\rm max}\approx2.4$ GeV.}.
In this work, NNLO matching was adopted, along with RGR up to the next-to-next-to-leading logarithmic order, validating the perturbative accuracy.
The authors accessed moments of the valence GPDs up to fifth order, including $A_{3,2}$, $A_{5,2}$, and $A_{5,4}$, with $z$-expansion and monopole-like parametrization of the $t$ dependence.

A parallel thread in SDF is the reconstruction of the $x$ dependence of partonic distributions.
The first study for zero-skewness GPDs was performed in Ref.~\cite{Bhattacharya:2024qpp} in the same setup as Ref.~\cite{Bhattacharya:2023ays}, but with additional high-statistics data for the largest nucleon boost, in the asymmetric frame.
The authors performed the matching at the level of ITDs and used a fitting ansatz to reconstruct $x$-dependent light-cone GPDs.
The latter were compared to results from LaMET, based on the same lattice data, concluding qualitative agreement, but quantitative differences in some regions of $x$, related to the different systematics in LaMET/SDF.
Nevertheless, these different systematics can actually be used as an advantage in a combined LaMET+SDF analysis, see the next subsection.

Another approach to the $x$ dependence reconstruction from SDF, including also reconstruction of the $t$ and $\xi$ dependence, was proposed in Ref.~\cite{Dutrieux:2026grg}\footnote{Lattice setup: $N_f=2+1$ clover fermions,  $a\approx0.094$ fm, $m_\pi\approx358$ MeV, $P_3^{\rm max}\approx2.7$ GeV.}. 
The authors considered the isovector unpolarized GPDs and extended the work of Ref.~\cite{HadStruc:2024rix}, applying Radyushkin's double-distribution formalism \cite{Radyushkin:1998es}.
Within this formalism, extracted GPDs automatically satisfy the crucial polynomiality property.
Since the lattice data are necessarily discrete and truncated, the authors used multidimensional Gaussian process regression to regularize the inverse problem and quantified uncertainties from possible model dependence on hyperparameters of the method.
In addition to the full reconstruction of the kinematic dependence of GPDs, an update of Mellin moments was presented, allowed by the extended coverage in nucleon boost, up to sixth order of both $A$- and $B$-type GFFs.

The final illustration of the application of the SDF approach to lattice data is Ref.~\cite{Cichy:2024afd}, which presented an exploratory study on integrating lattice-QCD results with experimental data for elastic scattering. 
This work was based on the same lattice data as Ref.~\cite{Bhattacharya:2024qpp} discussed above. 
The authors introduced a new double ratio observable (not to be confused with the double ratio used for renormalization of bare MEs), aimed at canceling some of the well-known lattice systematics.
A new type of a ``shadow'' term was also introduced in the extraction procedure, sensitive only to lattice-QCD results, which allowed for an investigation of the model dependence of the procedure.
The lattice data were then combined with experimental data on proton's and neutron's electric and magnetic form factors and the neutron charge radius.
As a result, the $t$-dependence of $A$- and $B$-type GFFs was obtained up to sixth order and tomographic pictures of the nucleon were presented.

\subsection{Next milestones: systematic effects, combining LaMET/SDF approaches, synergies of lattice extractions with experimental data}
All the work shortly discussed above demonstrates huge progress in lattice extractions of GPDs over a period of just a few years.
It also leads to some general conclusions for the way forward in the field.
An obvious conclusion is that several sources of systematic effects need yet to be addressed.
This concerns both typical lattice systematics (discretizations effects, finite volume effects, excited states contamination, or effects of non-physical quark masses) and other systematics that enter the extraction of light-cone GPDs.
The latter include effects in the renormalization of lattice-calculated MEs, the matching between Euclidean and Minkowski distributions (factorization) and the reconstruction of the continuous dependences on $x$, $t$, and $\xi$.
As a matter of fact, several of the works mentioned above have already addressed some of these systematics, in particular by employing the state-of-the-art hybrid renormalization and matching scheme, with leading-renormalon and renormalization-group resummations.

As detailed above, most of the lattice work concentrates on two approaches, LaMET and SDF, which utilize exactly the same lattice input of non-local MEs.
In Ref.~\cite{Bhattacharya:2024qpp}, the final $x$-dependent GPDs originating from the same lattice computation were shown, once analyzed within the LaMET framework and once within SDF.
The conclusion of quantitative differences stemming from different systematic effects is rather obvious and somewhat trivial, but it leads to the natural question whether LaMET and SDF can be combined in a single framework exploiting their complementary strengths.
The idea of such complementarity was put forward by Ji \cite{Ji:2022ezo} and illustrated with the valence PDF of the pion, where the LaMET extraction was aided by the second Mellin moment determined from SDF.
In Ref.~\cite{Chu:2025jsi}, a general unified framework was proposed to combine LaMET and SDF using artificial neural networks (ANNs) and applied to the extraction of nucleon's unpolarized PDFs and GPDs.
The essence of this approach is to use LaMET in its regime of applicability, i.e., in the region of $x$ between $x_{\rm min}\approx0.2$ and $x_{\rm max}\approx0.8$, and reconstruct the remaining $x$ region from SDF, using matched ITDs as constraints, i.e., without explicit extraction of Mellin moments.
Moreover, ANNs are used to regularize the inverse problem for both the $x$ and the $t$ dependence.
The authors showed that the unified framework is capable of reconstructing input model distributions from which mock data were generated in a way to mimic the discreteness and truncation of actual lattice data, translating reconstruction uncertertainty to final errors. 
The framework was also applied to the lattice GPDs data of Ref.~\cite{Bhattacharya:2024qpp}, yielding the full $x$ and $t$ dependence of zero-skewness $H$ and $E$ GPDs for valence quarks and tomographic pictures.
This work demonstrated that the extraction of light-cone PDFs and GPDs using the unified framework is more robust with respect to separate LaMET or SDF determinations and, thus, it is highly desirable that lattice analyses combine the two approaches rather than use them in isolation.
We view it as an important guideline for future work and one that ensures the most efficient use of the computationally most expensive part of lattice computation of non-local MEs.

Apart from unifying LaMET with SDF, a desired feature is to also exploit possible synergies of lattice calculations with experimental data. In the previous section, we discussed the combination of lattice GPD observables with elastic scattering data.
However, the synergy can go much further, leading to global analyses of lattice and experimental data.
The first framework aiming at this is the GUMP approach (GPDs through universal moment parametrization) \cite{Guo:2022upw,Guo:2023ahv,Guo:2024wxy,Guo:2025muf}, wherein a conformal moment space parametrization is used for a unified description across low- and high-$x$ regions at NLO perturbative accuracy.
The most recent and the most advanced from the point of view of exploiting lattice data global analysis of Ref.~\cite{Guo:2025muf} utilized experimental data on DVCS and deeply virtual $\rho$ meson production from Jefferson Lab
and the HERA accelerator at DESY, global fits of PDFs and electromagnetic form factors.
On the lattice side, asymmetric-frame data were used for unpolarized GPDs at $\xi=0$ ($x$-dependent LaMET extraction of Ref.~\cite{Bhattacharya:2022aob} and SDF-extracted moments of Ref.~\cite{Bhattacharya:2023ays} and $\xi\neq0$ (LaMET, Ref.~\cite{Chu:2025kew}), as well as for helicity GPDs (LaMET, Ref.~\cite{Bhattacharya:2023jsc}).
The outcome of the GUMP analysis was presented graphically in terms of exemplary tomographic pictures at selected values of $x$.

The GUMP framework provides the first large-scale demonstration of possible synergies between lattice, experiment and phenomenology. With the increasing amount of data on all sides, it becomes an important issue to facilitate data sharing between different collaborations. The first community effort in this direction is the creation of an open database of GPD-related data \cite{Burkert:2025gzu}. The goal of this database is to store both experimental and lattice-QCD data and aid global analyses.

\section{Conclusions and outlook}
In this chapter, we discussed the dynamical progress in calculations of GPDs from lattice QCD, concentrating on the two most popular approaches of quasi- and pseudo-distributions. While introduced in the context of PDFs, they were naturally extended to encompass off-forward kinematics of GPDs. Serious lattice work for GPDs started around 2019 and the period of just a few years witnessed tremendous progress in several theoretical and practical aspects:
\begin{itemize}
\item introduction of the formalism of asymmetric frames of reference, allowing for cost-efficient calculations of the full $t$ and $\xi$ dependence of GPDs, 
\item development of improved renormalization and matching schemes, significantly decreasing the corresponding uncertainties,
\item improved strategies for $x$ dependence reconstruction (neural networks, asymptotic extrapolations, Bayesian methods), regularizing the unavoidable inverse problem,
\item improved strategies for $t$ and $\xi$ dependence reconstruction (neural networks, double distributions), allowing for robust mapping of the full kinematic dependence and tomographic pictures,
\item proposed ways of merging information from LaMET and SDF, leading to the most efficient use of expensive lattice data,
\item proposed synergies of lattice and experimental data, enabling exploitation of their complementary strenghts.
\end{itemize}
Some of these aspects are common to PDFs and GPDs, but some are GPD-particular and focused on extracting the physical information specific to these generalized distributions.

The main challenges remaining fall into two categories:
\begin{itemize}
\item obtaining state-of-the-art lattice data that allow the quantification of all lattice-specific systematic uncertainties,
\item reliably translating the lattice data to final GPDs and related physical observables.
\end{itemize}

The first category requires performing calculations for several lattice ensembles, addressing discretization effects (extrapolating to the continuum limit), finite volume effects (large volumes needed also for good resolution in $t$), excited-states contamination (isolating the desired hadron's contribution), ideally at the physical pion mass to avoid chiral extrapolations. In addition, quark-disconnected contributions must be computed to achieve precise determinations of flavor-singlet and gluon GPDs, the latter yet to be explored numerically.
Moreover, the data needs to be obtained with sufficiently highly-boosted hadron states, which is the crucial prerequisite for the second category.
Both LaMET and SDF require large momenta for a reliable control of power corrections (genuine higher-twist effects, as well as hadron mass and $t$-dependent ones) appearing at the matching stage. The theoretical framework of the matching needs to be consistently extended to the NNLO order in perturbation theory for all physical cases of interest, with appropriate resummations for good perturbative control. Finally, methods for regularization of the inverse problem need to be further advanced and applied to the data, translating deficiencies of the data (their unavoidable discreteness and limited range) to uncertainties of the final objects.

While the above challenges are far from trivial to overcome, the present status of lattice GPDs is already advanced and it is not plagued by any problems of principle. Hence, the prospects of this field should be judged as very positive, with realistic hopes for extracting profound physical information about the internal structure of hadrons and achieving synergies with experiment and phenomenology. 

\section*{Acknowledgements}
We are grateful to the Editors of the Encyclopedia of Nuclear Physics for the invitation to write this contribution and their support. We also thank our collaborators and colleagues for insightful discussions on different topics related to this chapter, which shaped our views presented here.

K.~C.\ is supported by the National Science Centre (Poland) grant OPUS No.\ 2021/43/B/ST2/00497. 
M.C. acknowledges financial support from the U.S. Department of Energy, Office of Nuclear Physics,  under Grant No.\ DE-SC0025218.

\bibliographystyle{Harvard}
\bibliography{bibliography}

\end{document}